\documentclass[submit]{smj}

\usepackage{algorithm}
\usepackage[noend]{algpseudocode}

\usepackage{tikz}
\usetikzlibrary{fit}
\usetikzlibrary{calc}
\usetikzlibrary{decorations.pathreplacing}
\usetikzlibrary{arrows.meta}
\usetikzlibrary{positioning}

\Author{Sina Mews\Affil{1}\ORCID{0000-0003-1138-3185}, 
        Jan-Ole Koslik\Affil{1}\ORCID{0009-0004-1556-9053},
        and Roland Langrock\Affil{1}\ORCID{0000-0001-8206-6908}
}
\AuthorRunning{Sina Mews \textrm{et al.}}

\Affiliations{

\item Department of Business Administration and Economics,  
      Bielefeld University, 
      Bielefeld,
      Germany

}   

\CorrAddress{Sina Mews, 
             Department of Business Administration and Economics, 
             Bielefeld University,
             Universitätsstraße 25, 
             D--33615 Bielefeld, 
             Germany}
\CorrEmail{sina.mews@uni-bielefeld.de}
\CorrPhone{(+49)\;521\;106\;5103}

\Title{How to build your latent Markov model --- the role of time and space}
\TitleRunning{How to build your latent Markov model}

\Abstract{
Statistical models that involve latent Markovian state processes have become immensely popular tools for analysing time series and other sequential data. 
However, the plethora of model formulations, the inconsistent use of terminology, and the various inferential approaches and software packages can be overwhelming to practitioners, especially when they are new to this area. 
With this review-like paper, we thus aim to provide guidance for both statisticians and practitioners working with latent Markov models by offering a unifying view on what otherwise are often considered separate model classes, from hidden Markov models over state-space models to Markov-modulated Poisson processes. 
In particular, we provide a roadmap for identifying a suitable latent Markov model formulation given the data to be analysed. 
Furthermore, we emphasise that it is key to applied work with any of these model classes to understand how recursive techniques exploiting the models' dependence structure can be used for inference. 
The R package \texttt{LaMa} adapts this unified view and provides an easy-to-use framework for 
fast (C++ based) numerical maximum likelihood estimation 
of any of the models discussed in this paper, allowing users to tailor a latent Markov model to their data using a Lego-type approach. 
}

\Keywords{
continuous-time Markov chain; hidden Markov models; Markov-modulated Poisson processes; maximum likelihood; state-space models
}

\begin{document}

\maketitle


\section{Introduction}

Latent Markov models are stochastic process models for sequential data that are driven by latent Markovian system processes --- also referred to as, {\it inter alia}, state-space models, hidden Markov models, doubly stochastic processes, or dependent mixture models. 
Over the last two decades, these models 
have taken applied research by storm \citep{jackson2003, bartolucci2012latent, zucchini2016, mcclintock2020uncovering, bartolucci2022discrete}. This success story can be explained by their intuitive appeal, their mathematical tractability and versatility, and the various types of inference they allow for. 
Yet, while most empirical researchers are well-acquainted with the various flavours of regression, the same cannot be said for latent Markov models. 
As a consequence, researchers are likely to stick to one particular model class they are familiar with and shoehorn data sets to be analysed into that very model formulation, though it may be suboptimal or even inappropriate (cf.\ \citealp{hamaker2015modeling}). 
In this contribution, we thus aim to provide a roadmap to guide practitioners in the choice, the formulation, and the estimation of a suitable latent Markov model for their data.

The challenge of identifying a suitable model formulation for a given data set primarily concerns two choices to be made: whether to use a discrete-time or a continuous-time model formulation (see, e.g., \citealp{mcclintock2014, deHaan2017, loossens2021}), and whether to assume a discrete or a continuous state space (see, e.g., \citealp{domokos2004discrete, hamaker2015modeling, lavielle2018}). 
Classifying different model classes along these two dimensions, we provide an overview of what we regard as the most relevant classes of latent Markov models, namely discrete- and continuous-time hidden Markov models (HMMs, including Markov-switching regression models; \citealp{bartolucci2012latent,zucchini2016}), discrete- and continuous-time state-space models (SSMs; \citealp{hamilton1994state,auger2021}), and Markov-modulated Poisson processes (MMPPs; \citealp{fischer1993markov,ryden1994parameter}). 
We offer a unified view of these types of models (as, to some extent, was also done in, e.g., \citealp{zeng2013}, and \citealp{auger2021}) and argue that they can be regarded as one large family of models, which we summarise under the umbrella term ``latent Markov models''. 
In particular, the five model classes considered 
differ only in specific aspects of the model formulation, while sharing the same general structure and, importantly, allowing for the same inferential tools to be applied. 
Specifically, we focus on direct numerical likelihood maximisation and emphasise that the forward algorithm used for evaluating the likelihood can be applied for all model classes discussed. 
While the components of the forward algorithm differ slightly across the model classes, the inferential methods are effectively identical, thus unifying inference for latent Markov models. 
The new R \citep{r2024} package \texttt{LaMa} \citep{lama} exploits the latter point and offers a Lego-type toolbox for fast and convenient estimation of all model classes listed above.

The structure of the paper is as follows. We begin by introducing the five model classes considered --- namely discrete- and continuous-time HMMs and SSMs as well as MMPPs --- and discussing how to choose between them depending on the characteristics of the data to be analysed (Section~\ref{s:overview}). Subsequently, in Section~\ref{s:fitting}, we make our main point regarding the possible unification of these model classes, by showing that the structure of the likelihood is identical for each of the five models considered. In Section~\ref{s:appl}, we provide brief case studies for each of the five models considered --- these are supposed to help practitioners to develop an intuition for both the similarities but also the differences between the models.

\section{Overview of model classes and the choice thereof}
\label{s:overview}


We consider situations in which an observed data sequence $x_{t_1},x_{t_2},\ldots,x_{t_T}$
is assumed to depend on an underlying, non-observed state process that evolves stochastically over time.  
The observation times $t_1,\ldots,t_T$ may be regularly or irregularly spaced in time, and may in fact themselves be random variables --- as will be pointed out below, these different cases are tied to different (but mathematically similar) model formulations.

Such situations, with observed sequential data being driven by an unobserved state process, arise in various fields of application; for example: 
\begin{itemize}
\item observed animal movement depends on not directly observed behavioural modes (e.g., \citealp{michelot2016movehmm, beumer2020});
\item financial share returns depend on the underlying market volatility (e.g., \citealp{kim1998, bulla2006});
\item earthquake occurences and their magnitudes depend on the current seismic activity (e.g., \citealp{beyreuther2008, lu2012});
\item medical measurements such as biomarkers depend on the patient's underlying health state (e.g., \citealp{jackson2003, amoros2019});
\item survey/panel data on emotions and affect depend on latent psychological constructs and processes (e.g., \citealp{voelkle2012, wood2018}). 
\end{itemize}

When modelling such data that are driven by latent states, the first choice to be made is whether the model should operate in discrete or continuous time. 
This decision is usually dictated by the sampling scheme of the data at hand: 
for data collected at regular (i.e.\ equidistant) time intervals, discrete-time models whose results are interpreted with respect to the fixed sampling frequency can be used, while continuous-time models conveniently accommodate data collected irregularly or opportunistically over time. 
In the latter case, one further needs to distinguish whether observation times are non-informative or informative with respect to the underlying state process. 
If they are informative, the times at which observations are made are themselves random variables depending on the latent state process, leading to point process model formulations \citep{cox1980, daley2003}.

\begin{figure}[!tbh] 
    \centering
    \includegraphics[width=\textwidth]{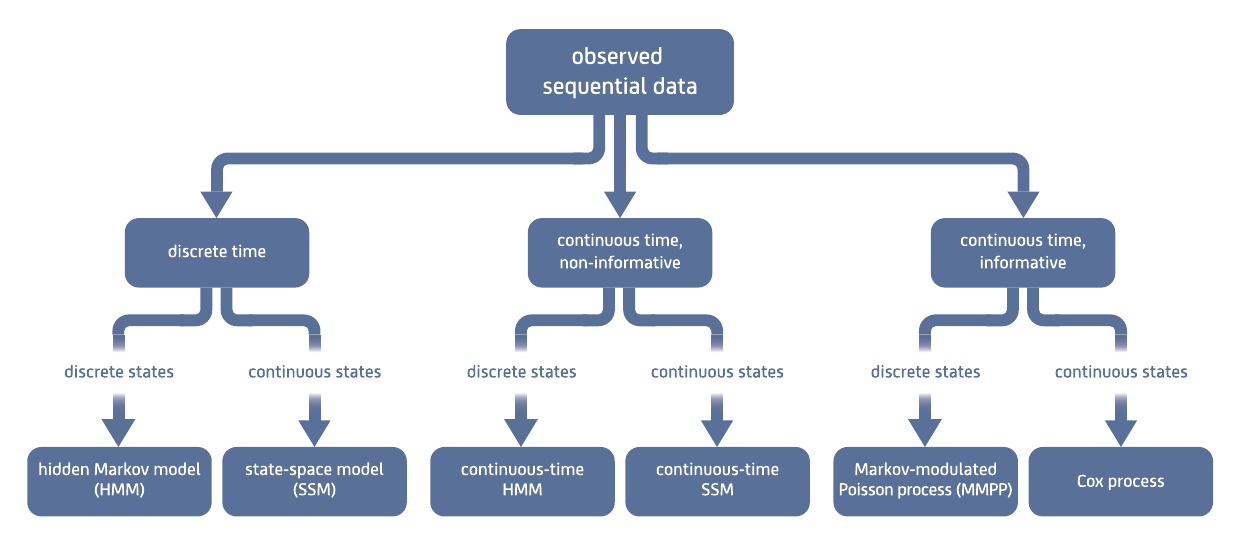}
    \vspace{-2em}
    \caption{Decision tree for identifying a suitable latent Markov model based on the role of time (i.e.\ the sampling scheme) and the nature of the state space (discrete vs.\ continuous states) 
    dictated or suggested by the data. 
    Regarding the former, models can be operationalised in discrete or continuous time, where observation times are either informative (i.e.\ they depend on the latent state process) or non-informative. }
    \label{fig:decisionTree}
\end{figure}

The second choice to be made concerns the nature of the state space, which is assumed to be either discrete or continuous. 
This decision depends on the (intended) interpretation of the latent states: 
for example, discrete states can correspond to physiological conditions like patients' disease states \citep{jackson2003} or animals being either alive or dead \citep{royle2008}, 
while continuous-valued states can for example correspond to real-valued locations in space \citep{jonsen2020} or to the gradually varying volatility of the financial market \citep{jacquier2002bayesian}. 
Summarising these modelling choices on the role of time and the state space leads to the classification of six popular latent Markov models shown in Figure~\ref{fig:decisionTree}. 
In this paper, however, we do not discuss Cox processes, as in terms of the associated inferential tools they do not fit into the unified view we are proposing. 
Note that we use the term hidden Markov model (HMM) to refer to the special case of a state-space model (SSM) with discrete-valued states.

\subsection{Basic model formulations}
\label{ss:modelFormulation}

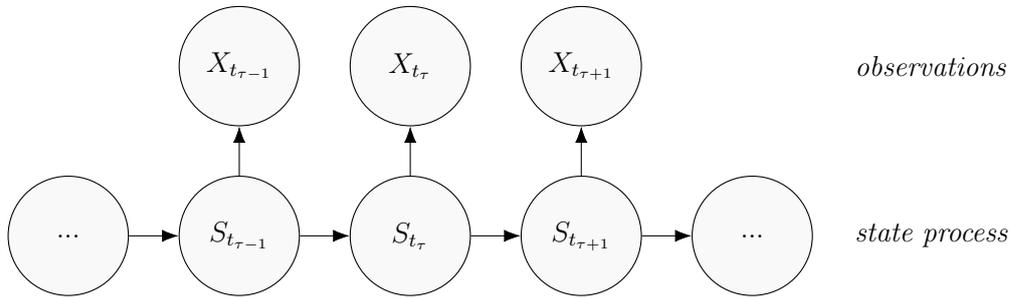
\begin{figure}[!thb]
    \centering
    \vspace{0.5cm}
    \begin{tikzpicture}[scale=0.9, transform shape]
    \node[circle,draw=black, fill=gray!5, inner sep=0pt, minimum size=50pt] (A1) at (0, -5) {...};
    \node[circle,draw=black, fill=gray!5, inner sep=0pt, minimum size=50pt] (A) at (2.5, -5) {$S_{t_{\tau-1}}$};
    \node[circle,draw=black, fill=gray!5, inner sep=0pt, minimum size=50pt] (B) at (5, -5) {$S_{t_\tau}$};
    \node[circle,draw=black, fill=gray!5, inner sep=0pt, minimum size=50pt] (C) at (7.5, -5) {$S_{t_{\tau+1}}$};
    \node[circle,draw=black, fill=gray!5, inner sep=0pt, minimum size=50pt] (C1) at (10, -5) {...};
    \node[circle,draw=black, fill=gray!5, inner sep=0pt, minimum size=50pt] (Y1) at (2.5, -2.5) {$X_{t_{\tau-1}}$};
    \node[circle,draw=black, fill=gray!5, inner sep=0pt, minimum size=50pt] (Y2) at (5, -2.5) {$X_{t_\tau}$};
    \node[circle,draw=black, fill=gray!5, inner sep=0pt, minimum size=50pt] (Y3) at (7.5, -2.5) {$X_{t_{\tau+1}}$};
    \node[text width=4cm,font=\itshape] at (13.5, -2.5) {observations};
    \node[text width=4cm,font=\itshape] at (13.5, -5) {state process};
    \draw[-{Latex[scale=1.3]}] (A1)--(A);
    \draw[-{Latex[scale=1.3]}] (A)--(B);
    \draw[-{Latex[scale=1.3]}] (B)--(C);
    \draw[-{Latex[scale=1.3]}] (C)--(C1);
    \draw[-{Latex[scale=1.3]}] (A)--(Y1);
    \draw[-{Latex[scale=1.3]}] (B)--(Y2);
    \draw[-{Latex[scale=1.3]}] (C)--(Y3);
    \end{tikzpicture}
    \caption{Dependence structure of basic latent Markov models:  the states follow a Markov process with first-order dependence, while the distribution of the observation at time $t_\tau, \tau = 1, \ldots, T,$ is completely determined by the active state $s_{t_\tau}$}.
    \label{fig:depStructure}
\end{figure}

Latent Markov models provide a flexible hierarchical framework by distinguishing the \textit{observation process} --- corresponding to the data sequence $\{X_t\}$ observed at time points $t \in \{ t_1, \ldots, t_T \}$ --- from the underlying latent \textit{state process} $\{S_t\}$. 
Note that we use uppercase letters of $X_t$ and $S_t$ to represent random variables, while lowercase letters correspond to realisations of the processes. 
In many applications, the observation process is regarded as a noisy version of the state process, with the latter typically (though not always) being of primary interest \citep{patterson2017statistical, mcclintock2020uncovering,auger2021}. 
In the most basic model formulations, the evolution of the states over time is modelled as a Markov process, while the distribution of the observation at time $t \in \{ t_1, \ldots, t_T \}$ is completely determined by the underlying state $s_{t}$, as illustrated in Figure~\ref{fig:depStructure}. 
In all model classes considered below, the observation process is therefore fully specified by the conditional distribution $f(x_t \mid s_t)$, 
where $f$ denotes either a density or a probability mass function. 
For continuous-time models, this implies that the snapshot property needs to be satisfied, meaning that the observation is only allowed to depend on the underlying state at that time instant (such as blood biomarkers depending only on the health status at the time of the measurement) and not any previous state in a time interval prior to the observation (such as the distance travelled by an animal depending on the trajectory of the continuous-time behavioural state process); see \citet{glennie2022} for more details on the snapshot property and its violations. 
In the following, we present each model class in its most basic model formulation, briefly discussing typical extensions below.

\textbf{Hidden Markov models (HMMs)} 
involve discrete-valued states $S_t \in \{1, \ldots, N\}$ evolving in discrete time, such that $t = t_1, \ldots, t_T$ simplifies to $t = 1, \ldots, T$ with equidistant time steps. 
Crucially, data featuring neither equidistant time steps nor any other meaningful regular sampling unit should not directly be modelled using (discrete-time) HMMs but would need to be forced into a regular time grid (e.g., using data transformations or by including missing values), as the model parameters of the state process are interpreted with respect to 
the fixed interval length between observations (cf.\ Section~\ref{s:discussChoices}).
The latent state process is modelled as a (discrete-time) Markov chain characterised by the initial state distribution 
$$ \boldsymbol{\delta}^{(1)} = (\delta_1^{(1)}, \ldots, \delta_N^{(1)}) = \bigl(\text{Pr}(S_1 = 1), \ldots, \text{Pr}(S_1 = N) \bigr) $$  
and the (homogeneous) state transition probabilities 
$$\gamma_{ij} = \text{Pr}(S_t = j \mid S_{t - 1} = i), \quad i, j = 1, \ldots, N.$$ 
The latter are summarised in the $N \times N$ transition probability matrix (t.p.m.)
$\boldsymbol{\Gamma} = (\gamma_{ij})_{i, j = 1, \ldots, N}$, 
where the $i$th row contains the conditional one-step-ahead distribution of the state process given that the current state is $i$, and hence is constrained to sum to $1$. 
Conditional on the states, the observations are generated by one of $N$ possible emission distributions $f_j(x_t) = f(x_t \mid s_t = j)$, $j = 1, \ldots, N,$ which can be continuous (e.g., normal or gamma) or discrete (e.g., Poisson or binomial), and also multivariate in case of observation vectors $\mathbf{x}_t$. When some parameters, typically the mean, of these emission distributions are modelled as additionally depending on covariates --- which does not complicate inference except for a potential increase in numerical instability --- then these models are typically referred to as Markov-switching (or regime-switching) regression models (\citealp{kim2008estimation,hamilton2010regime}; and see, e.g., \citealp{bartolucci2014latent} for a review on including covariates into HMMs). 
For a comprehensive introduction to HMMs, see, e.g., \citet{zucchini2016}.

In \textbf{state-space models (SSMs)}, 
the state process also operates in discrete time, but the state space is assumed to be continuous, hence allowing for more gradual changes. 
Such a continuous-valued state process $S_t \in \mathbb{R}$, $t = 1, \ldots, T$, is often modelled using an autoregressive  (AR) process of order 1, where
\begin{equation}
    S_t = \phi(S_{t-1} - \mu) + \mu + \sigma \epsilon_t, \quad \epsilon_t \stackrel{\text{iid}}{\sim} \mathcal{N}(0, 1), 
\label{eq:AR1}
\end{equation}
with persistence parameter $-1 < \phi < 1$, long-term mean $\mu \in \mathbb{R}$, and standard deviation $\sigma > 0$ of the error process. 
The emission distribution of observation $x_t$ can be either discrete or continuous and is again fully determined by the underlying state $s_t$. 
For comprehensive references on SSMs with a view on applications, we refer to \citet{durbin2012} and \citet{zeng2013} for the field of economics and finance, or \citet{newman2014modelling} and \citet{auger2021} for ecology.

Analogously to HMMs, the underlying state process of \textbf{continuous-time hidden Markov models (ctHMMs)} involves $N$ discrete-valued states. 
However, the state process is modelled as evolving in continuous time, which allows the accommodation of (possibly) irregularly sampled data at times $t = t_1, \ldots, t_T$. 
In these models, the state process is governed by a continuous-time Markov chain with a transition intensity matrix (also called infinitesimal generator matrix)
$$\mathbf{Q} = (q_{ij})_{i, j = 1, \ldots, N},$$ 
where the off-diagonal elements $q_{ij} \geq 0$ are defined as 
$$q_{ij}=\lim\limits_{\triangle t \to 0}\frac{\text{Pr}(S_{t+\triangle t}=j \mid S_t=i)}{\triangle t}, $$ 
and thus can be interpreted as the rates at which transitions from state $i$ to state $j$ occur.
The diagonal elements are $q_{ii} = -\sum_{j \neq i} q_{ij}$, such that the rows sum to $0$. 
Based on the intensity matrix $\mathbf{Q}$, we can derive the t.p.m.\ $\boldsymbol{\Gamma}(\Delta_\tau)$ for a given time interval $\Delta_\tau = t_\tau - t_{\tau-1}, \tau = 2, \ldots, T,$ as the matrix exponential 
\begin{equation*}
    \boldsymbol{\Gamma}(\Delta_\tau) = \text{exp}(\mathbf{Q} \Delta_\tau) = \sum_{d=0}^\infty \frac{1}{d!} \mathbf{Q}^d \Delta_\tau^d. 
\end{equation*}
The other model components, i.e.\ the initial state distribution $\boldsymbol{\delta}^{(1)}$ and the emission distributions $f_j(x_t)$, are the same as in HMMs, provided that the snapshot property holds. 
See, for example, \citet{dobrow2016introduction} for an introduction to continuous-time Markov chains and \citet{jackson2003} for ctHMMs.

In \textbf{continuous-time state-space models (ctSSMs)},
the continuous-valued states evolve in continuous time.
To model such a state process $\{S_t\}_{t \geq 0}$, stochastic differential equations (SDEs) can be used. 
In principle, any SDE can be chosen, but inference is much simpler for SDEs whose transition densities have an explicit analytical form. 
One prominent example is the Ornstein-Uhlenbeck (OU) process
\begin{equation}
    d S_t = \theta (\mu - S_t) dt + \sigma d W_t, 
\label{eq:OUprocess}
\end{equation}
where $\theta$ controls the strength of reversion to the long-term mean $\mu \in \mathbb{R}$, $\sigma > 0$ controls the strength of fluctuations, and $W_t$ denotes the Wiener process (also called Brownian motion); see for example \citet{iacus2008simulation} and \citet{maller2009ornstein} for more details. 
As in SSMs, the conditional distribution of observation $x_t$ given the underlying state $s_t$ can be either discrete or continuous. 
For introductory texts on ctSSMs, we refer to the aforementioned references on SSMs, most of which treat both discrete- and continuous-time models.

\textbf{Markov-modulated marked Poisson processes (MMMPPs)} 
are suited to situations in which the observation times themselves are informative with respect to the state process, i.e.\ when the underlying state affects the chance of observations being made, such that the mere presence of observations tells us something about the underlying state (for example, an animal is more likely to be caught by a camera trap when active, and a patient is more likely to be seen by a doctor when ill).  
In MMMPPs, the observed time points follow a Poisson arrival process, whose event rate $\lambda_j, j = 1, \ldots, N$, is selected by the state of the underlying continuous-time $N$-state Markov chain. 
Consequently, if the unobserved state $j \in \{ 1,\ldots, N \}$ remains constant within some time interval $\Delta$, then the number of events in that time interval is Poisson distributed with mean $\lambda_j \Delta$, and the waiting times between consecutive events, $y_\tau = t_\tau - t_{\tau - 1}, \tau = 2, \ldots, T,$ are exponentially distributed with parameter $\lambda_j$. 
Similar to ctHMMs, the state process is governed by the transition intensity matrix $\boldsymbol{Q}$, while the observations (also called \textit{marks}) are determined by the emission distributions $f_j(x_t)$ (assuming that the snapshot property is fulfilled). 
Overall, the state process hence governs both the observation process and the observation times. 
If only the observation times are modelled without additional data collected at these time points, i.e.\ without marks, the model simplifies to a Markov-modulated Poisson process (MMPP) comprising only the state process and the event rates. 
General references on MMPPs are, e.g., \citet{fischer1993markov} and \citet{ryden1994parameter}, while MMMPPs are described, e.g., in \citet{lu2012} and \citet{mews2023markov}.

These basic model formulations can and often need to be extended in various ways, for example by allowing for additional dependencies in the observation or the state process (e.g., semi-Markov state processes or additional autoregressive components in the observation process; \citealp{langrock2012flexible,maruotti2019hidden,lawler2019conditionally}), or by including random effects \citep{bartolucci2011assessment}, or, importantly, covariates in the model \citep{kim2008estimation,maruotti2011mixed,whittington2022towns}. 
In many applications, the inclusion of covariates is particularly relevant for addressing the empirical research questions, for example when the aim is to investigate a system's response to internal and external drivers (e.g., an animal's behavioural dynamics in response to anthropogenic activities, a patient's disease dynamics in response to medical treatment, or the economic market's response to fiscal policy changes). 
In principle, for each of the model classes presented above, any model component can be specified as a function of covariates. We will give several such examples in the case studies in Section~\ref{s:appl}.

\subsection{Discussion of modelling choices}
\label{s:discussChoices}

Although choosing between the different model classes may appear straightforward, in practice it is often not. 
Regarding the operationalisation of time, regularly sampled data can be analysed using both discrete- and continuous-time models. 
In particular, the latter contain the information needed to reconstruct corresponding discrete-time parameters, and offer the advantage of being independent of the temporal resolution at which the data were collected, thus enabling model comparisons across studies with different sampling designs. 
From an applied perspective, however, many researchers perceive discrete-time models to be simpler to formulate, fit, and interpret than their continuous-time counterparts, which are often mathematically more complex and challenging (e.g., regarding statistical inference; cf.\ \citealp{mcclintock2014, ryan2018}). 
As a consequence, discrete-time models usually constitute the default choice for data collected at regular time intervals.  

For data sampled irregularly in time, the situation is more complicated. 
Conceptually, such data can only be analysed in a meaningful way using continuous-time models that account for varying lengths of time intervals between observations, but data transformations can, in principle, be used to force the observations onto a regular sampling grid. 
To this end, the data are typically either subsampled, interpolated, or aggregated to derive equally spaced observations in time and then apply discrete-time methods \citep{mcclintock2017incorporating,rushing2023ecologist}. 
Such a two-stage approach comes with several drawbacks.
First, choosing an arbitrary regular interval length between observations introduces subjectivity into the analysis, which is particularly relevant as the resulting inference is tied to this specific temporal resolution. 
Second, a discrete-time approach based on thinned or aggregated data discards information inherent in the raw observations by neglecting the exact observation times, while approaches based on interpolated data involve approximation errors and possibly a lack of uncertainty propagation. 
As a consequence, such data transformations not only add uncertainty to the estimated model and its corresponding conclusions but can ultimately lead to biased estimates \citep{schlagel2016, deHaan2017, kuiper2018drawing}. 
Consequently, continuous-time model formulations such as ctHMMs or ctSSM should generally be preferred for modelling irregularly sampled sequential data.

Another high-level decision concerning the operationalisation of time --- which is often overlooked \citep{farzanfar2017} --- is the potential informativeness of observation times. 
Such informative processes emerge for example in health care studies or medicine if longitudinal observations are collected at patient-initiated visit times to physicians, likely indicating sickness \citep[e.g.][]{lange2015, mews2023markov}. 
When neglecting such correlations between the (frequency of) observation times and the latent state process (e.g., corresponding to a patient's disease severity), parameter estimates are potentially biased \citep{pullenayegum2016, gasparini2020}. 
Thus, it is crucial to account for informative observation times providing information on the latent state process, which can be modelled using doubly stochastic point processes \citep{grandell1976doubly, cox1980} such as MM(M)PPs. 

Regarding the nature of the state space, deciding between discrete- or continuous-valued states is only straightforward for variables that correspond to genuine entities observed partially or with errors (e.g., geographical locations or the presence/absence of diseases).
In contrast, hypothetical constructs in the latent state process can translate to either discrete- or continuous-valued states essentially depending on the researcher's preference, with prominent examples being the nervousness of the financial market \citep{kim1998, bulla2006} or a player's hotness in sports \citep{wetzels2016, otting2020hot}. 
Considering mathematical aspects, models with discrete state spaces are (usually) simpler to fit and often easier to interpret, provided that the number of states is relatively small. 
For models with a large number of states, the number of parameters increases rapidly, while interpretability diminishes. 
Models with continuous state spaces are usually more parsimonious in terms of the number of parameters but are generally less mathematically tractable and in many applications perceived to be less interpretable. 
Overall, the nature of the state space is a key conceptual decision by the modeller, which should primarily be guided by the following question:  
Does the entity represented by the state process change rather gradually (i.e.\ continuously) over time or does it jump between distinct (discrete) states? The answer to this question is often straightforward. When it is not, then model selection criteria can be consulted, although their application in the context of latent Markov modelling comes with its own caveats \citep{pohle2017selecting}. 

In particular, for latent Markov models with discrete states, practitioners face the crucial but often difficult task of order selection (i.e.\ determining the number of hidden states). 
As the role of states is data-driven, they can capture complex structures neglected in the model formulation --- like outliers, multimodality, and individual heterogeneity featured in many real data sets. 
Therefore, including additional states in the model frequently leads to an improvement as measured by model selection criteria, even if these additional states may not allow for any meaningful interpretation. 
Information criteria are nevertheless widely used because they are simple to calculate and can provide some guidance on model selection (see, e.g., \citealp{bacci2014}, \citealp{pohle2017selecting}, and \citealp{buckby2023}, for comprehensive discussions of using information criteria for latent Markov models).   
Despite efforts to facilitate order selection in models with latent discrete states --- for example, using cross-validated likelihood scores \citep{celeux2008}, quasi-likelihood ratio tests \citep{holzmann2016}, or reversible jump Markov chain Monte Carlo (MCMC) procedures \citep{luo2021} --- choosing an adequate number of states remains challenging. 
We thus advise taking into account both empirical evidence and more practical considerations --- such as computational costs, expert knowledge, and the specific research question tackled --- when determining a discrete number of latent states. 

Due to the different modelling choices and options --- in particular, possible data transformations, time and state discretisations, and (possibly subjective) assumptions about the (non-)informativeness of observations and the nature of the state space --- researchers can to some extent force any data set into any of the model classes discussed, typically the one they happen to know best or feel most comfortable with. 
In many cases, this amounts to simplifications of the time formulation and state process, as models in discrete time and those with discrete-valued states are mathematically less complex and perceived as easier to understand and interpret. 
However, given the strong similarities between the various model classes --- especially with respect to inferential tools as detailed in the next section --- there is no need to limit oneself to building expertise on and working with only one particular latent Markov model.

\section{Statistical inference for latent Markov models}
\label{s:fitting}

Parameter estimation for latent Markov models is made difficult by the fact that the underlying states are not observed. 
The three main approaches to conduct estimation are Markov chain Monte Carlo (MCMC; \citealp{brooks2011}), likelihood maximisation via the expectation-maximisation (EM) algorithm \citep{baum1970maximization, dempster1977}, and direct numerical likelihood maximisation \citep{myung2003}, with none of these being generally superior to the others \citep{ryden2008versus, macdonald2014numerical, zucchini2016}. 
Specifically, the Bayesian framework, i.e., the use of MCMC, can prove advantageous for random effects modelling, and directly delivers a comprehensive uncertainty quantification via the posterior distribution of the model parameters.  
On the downside, for HMMs the approach is known to be usually much slower to converge than the frequentist alternatives, model selection is usually less straightforward, and technical issues such as label switching are not always trivial to address (cf.\ \citealp{ryden2008versus, zucchini2016}). 
The EM algorithm is conceptually very appealing, is typically relatively robust with respect to the choice of its initialisation and can be especially powerful in scenarios where the separation into state-specific optimisation problems leads to a substantial simplification (see, e.g., \citealp{bartolucci2014latent,langrock2018spline, feldmann2023}). However, EM can also be slow to converge, and generally requires the tailoring of the algorithm to the given problem and associated model formulation, which can be tedious in practice. 
Direct numerical likelihood maximisation is easy to understand and implement, is usually very fast and easy to work with, and, perhaps most importantly, usually requires only minimal modifications when varying the model formulation. 
On the downside, the ever-increasing complexity of latent Markov models being used in real-data analyses comes with an increase also of the numerical instability, which can be more difficult to address and handle than with the alternative approaches.
In particular, within the Bayesian paradigm the incorporation of prior information can help to control the navigation through the parameter space.
For a comprehensive comparison of Bayesian and frequentist fitting methods, we refer to \citet{auger2021} and references therein. We would like to stress that any of these approaches is perfectly valid and practitioners should generally resort to whatever approach they feel most comfortable with. 

In this contribution, we focus on the third approach, using standard off-the-shelf numerical optimisers to maximise the likelihood function, which is evaluated using the forward algorithm. 
The key strength of this approach is that the forward algorithm requires only minimal adjustments across the different model classes, meaning that mastering this recursive technique is key for applied work. In particular, provided a good understanding of the forward algorithm, direct numerical likelihood maximisation can fairly easily be implemented to conduct parameter estimation in any of the model classes considered.

\subsection{Likelihood evaluation using the forward algorithm} 
\label{s:inference}

The general idea of the forward algorithm is to traverse along the sequence of observations $(x_{t_1}, x_{t_2}, \ldots, x_{t_T})$ and update the likelihood at each time step, while retaining probabilistic information on the state currently active (cf.\ \citealp{zucchini2016}). 
For latent Markov models with continuous-valued states, the state space is finely discretised to (arbitrarily accurately) approximate the likelihood and render the forward algorithm applicable \citep{langrock2011, mews2024}. 
Evaluating the (approximate) likelihood with respect to the unknown parameter vector $\boldsymbol{\theta}$ then amounts to evaluating the following matrix product:
\begin{equation}
    \mathcal{L}(\boldsymbol{\theta}) = \boldsymbol{\delta}^{(1)} \mathbf{P}(x_{t_1}) \boldsymbol{\Omega}^{(2)} \mathbf{P}(x_{t_2}) \boldsymbol{\Omega}^{(3)} \cdot \dotsc \cdot \boldsymbol{\Omega}^{(T)} \mathbf{P}(x_{t_T}) \boldsymbol{1},
\label{eq:llk}
\end{equation}
where $\boldsymbol{\delta}^{(1)}$ is the initial state distribution, $\boldsymbol{1}$ is a column vector of ones, $\mathbf{P}(x_{t_\tau}), \tau = 1, \ldots, T$, is a diagonal matrix containing the conditional distributions of $x_{t_\tau}$ given $s_{t_\tau}$, and $\boldsymbol{\Omega}^{(\tau)}$ is a transition operator, specifically a matrix containing the probabilities of switching between the states within the time interval $[t_{\tau - 1}, t_\tau]$.
Note that $\boldsymbol{\delta}^{(1)}$ can be either estimated or assumed to be the stationary distribution $\boldsymbol{\delta}$ of the state process, if it exists. 
Expression (\ref{eq:llk}) applies to all model classes discussed in Section~\ref{ss:modelFormulation}, with only the components, especially $\boldsymbol{\Omega}^{(\tau)}$, differing across models, as discussed below.

\textbf{HMMs}: 
Here the $N$-state Markov chain operates in discrete time assuming fixed interval lengths between consecutive observations, with respect to which the state transition probabilities, summarised in the t.p.m.\ $\boldsymbol{\Gamma}$, are interpreted.
Therefore, $$ \boldsymbol{\Omega}^{(\tau)} = \boldsymbol{\Gamma}, \quad \text{with} \quad \gamma_{ij} = \text{Pr}(s_t = j  \mid s_{t-1} = i), \quad i, j = 1, \ldots, N, $$ 
if the state process is homogeneous, and $\boldsymbol{\Omega}^{(\tau)} = \boldsymbol{\Gamma}^{(t)}$, $t = 1, \ldots, T$, if it depends on time-varying covariates (see the example application in Section~\ref{s:appl:HMMs}). 
For the initial distribution --- if it is not estimated --- the stationary distribution $\boldsymbol{\delta}$ of a homogeneous Markov chain can be obtained by solving 
$ \boldsymbol{\delta} \boldsymbol{\Gamma} = \boldsymbol{\delta}$ subject to $\sum_{j = 1}^N \delta_j = 1$. The densities or probability mass functions of the emission distributions are summarised in   
$\mathbf{P}(x_t) = \text{diag} \bigl(f_1(x_t), \ldots, f_N(x_t) \bigr)$.

\textbf{SSMs}: 
The likelihood of an SSM involves the integration over all possible values of the state process at each observation time. 
For general SSMs with possibly non-linear and non-Gaussian processes, this multiple integral cannot be evaluated analytically. 
However, numerical integration via a fine discretisation of the state space can be used to arbitrarily accurately approximate the likelihood. 
This discretisation effectively amounts to an approximation of the SSM by an HMM with many states and a structured t.p.m.\ (\citealp{langrock2011,zucchini2016}), hence paving the way to the application of the forward algorithm.  
Specifically, we divide the possible range of states into $m$ intervals $B_i = (b_{i-1}, b_i)$, $i = 1, \ldots, m$, of equal length $h = (b_m - b_0) / m$ and let $b_i^*$ denote the $i$th interval's midpoint. 
Then $$ \boldsymbol{\Omega}^{(\tau)} = \boldsymbol{\Gamma}, \quad \text{with} \quad \gamma_{ij} = h \cdot f(s_t = b_j^* \mid s_{t-1} = b_i^*), \quad i, j = 1, \ldots, m, $$ 
for a homogeneous, discrete-time state process with regular observation times $t = 1, \ldots, T$. 
In this approximation, $\gamma_{ij}$ is approximately equal to the probability that $s_t$ falls into the interval $B_j$, given that $s_{t-1}$ was in the interval $B_i$. 
If, for example, the states follow an AR(1) process with persistence $\phi$, long-term mean $\mu$, and error variance $\sigma^2$ as in Equation~(\ref{eq:AR1}), then 
$$ S_t \mid S_{t-1} = b_i^* \sim \mathcal{N} \left(\phi (b_i^*-\mu)+\mu, \sigma^2 \right), $$
such that $\gamma_{ij}= h \cdot f_{\mathcal{N} \left(\phi (b_i^*-\mu)+\mu,\, \sigma^2 \right)}(b_j^*)$.
In this example, we can further derive $\boldsymbol{\delta}^{(1)}$ based on the stationary distribution of the AR(1) process, which is a normal distribution with mean $\mu$ and variance $\sigma^2/(1-\phi^2)$. 
The final ingredient to the likelihood is the $m \times m$ matrix $\mathbf{P}(x_t) = \text{diag} \bigl(f(x_t \mid s_t = b_1^*), \ldots, f(x_t \mid s_t = b_m^*) \bigr)$. 
The likelihood approximation can be made arbitrarily accurate by increasing the number of intervals $m$ (see \citealp{langrock2011}, for details on how to specify $m$ as well as the state boundaries $[b_0, b_m]$), but is feasible only in case of low-dimensional state spaces.

\textbf{ctHMMs}: 
When the discrete-valued state process evolves in continuous time, then it is governed by the $N \times N$ transition intensity matrix $\mathbf{Q}$, allowing for state switches at any point in time. 
For an interval of length $\Delta_\tau = t_\tau - t_{\tau-1}$, the probabilities of switching states from time $t_{\tau-1}$ to time $t_\tau$ are then obtained as the matrix exponential  
$$ \boldsymbol{\Omega}^{(\tau)} = \boldsymbol{\Gamma}(\Delta_\tau) = \text{exp}(\mathbf{Q} \Delta_\tau). $$
While the  $N \times N$ matrix $\mathbf{P}(x_{t_\tau}) = \text{diag} \bigl(f_1(x_{t_\tau}), \ldots, f_N(x_{t_\tau}) \bigr)$ is the same as for HMMs (but requiring the snapshot property; cf.\ \citealp{glennie2022}), the stationary distribution $\boldsymbol{\delta}$ of a homogeneous continuous-time Markov chain is obtained as the solution to 
$ \boldsymbol{\delta} \mathbf{Q} = \mathbf{0}$ subject to $\sum_{j = 1}^N \delta_j = 1$.

\textbf{ctSSMs}: 
Analogously, as for SSMs, we can discretise the state space and approximate the transition probabilities by focusing on the intervals' midpoints, but now additionally account for the length of time between consecutive observations:
$$ \boldsymbol{\Omega}^{(\tau)} = \boldsymbol{\Gamma}(\Delta_\tau), \quad \text{with} \quad \gamma_{ij}(\Delta_\tau) = h \cdot f(s_{t_\tau} = b_j^* \mid s_{t_{\tau-1}} = b_i^*), \quad i, j = 1, \ldots, m, $$
with (possibly irregularly spaced) observation times $t_\tau$.
If, for example, the state process is modelled as an OU process (cf.\ Equation~\ref{eq:OUprocess}), then it has the Gaussian transition density 
$$ S_{t_\tau} \mid S_{t_{\tau-1}} = b_i^* \sim \mathcal{N} \Bigl( \text{e}^{-\theta \Delta_\tau} b_i^* + \mu \bigl(1- \text{e}^{-\theta \Delta_\tau}\bigr), \quad 
    \frac{\sigma^2}{2\theta} \bigl(1- \text{e}^{-2\theta \Delta_\tau}\bigr) \Bigr), $$
based on which the transition probabilities $\gamma_{ij}(\Delta_\tau)$ can easily be obtained. 
The stationary distribution of the OU process, an $\mathcal{N} \bigl(\mu, 0.5 \sigma^2/\theta \bigr)$ distribution, can be used to derive $\boldsymbol{\delta}^{(1)}$. 
The $m \times m$ matrix $\mathbf{P}(x_t) = \text{diag} \bigl(f(x_t \mid s_t = b_1^*), \ldots, f(x_t \mid s_t = b_m^*) \bigr)$ has the same structure as for SSMs (but like ctHMMs, requires the snapshot property). 
Additional details on statistical inference in ctSSMs using the forward algorithm can be found in \cite{mews2024}.

\textbf{MMMPPs}: 
Assuming the times at which observations are made to be random and also driven by the latent $N$-state continuous-time Markov chain, we obtain 
$$ \boldsymbol{\Omega}^{(\tau)} = \exp \bigl( (\mathbf{Q} - \boldsymbol{\Lambda}) y_\tau \bigr) \boldsymbol{\Lambda}, $$
where $\boldsymbol{\Lambda} = \text{diag}(\lambda_1, \ldots, \lambda_N)$ is a diagonal matrix containing the state-dependent Poisson event rates, and $y_\tau = t_\tau - t_{\tau - 1}$, $\tau = 2, \ldots, T,$ are the waiting times between consecutive observations. For the latter, we use a different notation than for \mbox{ctHMMs} and ctSSMs to emphasise that these are now random variables. 
For very small $y_\tau$, by the Lie product formula, the matrix $\boldsymbol{\Omega}^{(\tau)}$ can be understood as the product of the probabilities a) of transitioning between the different states, represented by $\exp ( \mathbf{Q} y_\tau )$, b) of no event occurring within the interval of length $y_\tau$, represented by $\exp ( - \boldsymbol{\Lambda} y_\tau )$, and c) of observing an event in the infinitesimal interval at the end of the waiting time $y_\tau$, represented by $\boldsymbol{\Lambda}$. Formally, $\boldsymbol{\Omega}^{(\tau)}$ results as the solution to the Kolmogorov forward equations of the joint process comprising the Markov chain and the process counting the number of arrivals \citep[see][]{meier1987fitting}.
In contrast to the other model classes, $\boldsymbol{\Omega}^{(\tau)}$ thus does not correspond to a transition probability matrix, 
nevertheless the likelihood structure remains identical to Equation~(\ref{eq:llk}), rendering the forward algorithm applicable. 
The initial distribution $\boldsymbol{\delta}^{(1)}$ is the same as for ctHMMs. 
The same holds for the matrix $\mathbf{P}(x_{t_\tau})$ containing the conditional distribution of the observations, i.e.\ the marks. 
In the simpler special case without marks, i.e.\ an MMPP, we instead set $\mathbf{P}(x_{t_\tau}) = \mathbf{I}$. 

Once any of the above models has been chosen as a candidate to be fitted to given data, the corresponding likelihood as given in (\ref{eq:llk}) with its model-specific components can easily be implemented and numerically maximised using standard optimisation routines such as Newton-Raphson methods (subject to well-known technical issues, as comprehensively discussed in \citealp{zucchini2016}). The (log-)likelihood evaluation itself typically requires only a few lines of code, as illustrated by the pseudo-code below. Four of these lines of code (coloured in grey) are required to avoid numerical underflow (cf.\ Chapter 3 in  \citealp{zucchini2016}) 

\begin{algorithm}
\caption{Scaled forward algorithm to compute the log-likelihood}
\label{algo:scaled}
\begin{algorithmic}

\Require $\boldsymbol{\delta}^{(1)}$, $\{ \boldsymbol{\Omega}^{(\tau)} \}_{\tau=2, \dotsc, T}$, $\{ \textbf{P}(x_{t_{\tau}}) \}_{\tau = 1, \dotsc, T}$
\State calculate $\text{\textbf{foo}} = \boldsymbol{\delta}^{(1)} \textbf{P}(x_{t_1})$
{\color{gray}
\State set $\ell = \log(\text{sum(\textbf{foo})})$
\State set $\boldsymbol{\phi} = \text{\textbf{foo}} / \text{sum(\textbf{foo})} $ 
}

\For{$\tau = 2, \ldots, T$}
    \State calculate $\text{\textbf{foo}} = \boldsymbol{\phi} \boldsymbol{\Omega}^{(\tau)} \textbf{P}(x_{t_{\tau}})$ 
    {\color{gray}
    \State update $\ell = \ell + \log(\text{sum(\textbf{foo})})$
    \State set $\boldsymbol{\phi} = \text{\textbf{foo}} / \text{sum(\textbf{foo})}$ 
    }
\EndFor

\State \Return \(\ell\)

\end{algorithmic}
\end{algorithm}

To further simplify this process and, importantly, help users minimise the computational cost of such an implementation, we 
created the R package \texttt{LaMa} \citep{lama}, which allows for very fast and convenient estimation of all model classes discussed. In \texttt{LaMa}, the most time-consuming part of the parameter estimation, namely the (repeated) evaluation of the likelihood function via the forward algorithm, is outsourced to C++, which drastically improves the computational speed. 
Users merely need to specify the three likelihood ingredients, i.e.\ the initial distribution $\boldsymbol{\delta}^{(1)}$, the transition operator $\boldsymbol{\Omega}^{(\tau)}$ and the emission distributions contained within $\mathbf{P}(x_{t_{\tau}})$.
Additionally, the package contains several helper functions for the various special cases of $\boldsymbol{\Omega}^{(\tau)}$, allowing for convenient and fast computation for example of inhomogeneous t.p.m.s or the matrix exponential.
Moreover, \texttt{LaMa} is fully compatible with automatic differentiation as implemented in the new R package \texttt{RTMB} \citep{kristensen2024rtmb}. The ability to evaluate gradients at the same computational cost as the log-likelihood function drastically reduces the overall computation time, and accurate gradient information allows for much higher precision. This renders direct numerical likelihood maximisation even more attractive compared to the alternative approaches, mainly MCMC and the EM algorithm.

Alternatively, various other R packages exist that can be used to simulate and estimate latent Markov models. 
Many of these packages are motivated by specific data applications, for example for modelling animal movement data (e.g., \citealp{mcclintock2018momentuhmm}) or disease progression based on panel data (e.g., \citealp{jackson2011msm}). 
A comprehensive list of available R packages for HMMs can be found in \citet[][Table~2]{mcclintock2020uncovering} --- except for, {\it inter alia}, the recent R packages by \citet{fHMM} and \citet{michelot2022hmmtmb} --- while \citet[][Table~2]{auger2021} provide a list of possible R packages for fitting SSMs to data. 
In contrast, only a few R packages currently include methods for simulating and estimating MMPPs (i.e.\ only without marks), including \texttt{ppdiag} (\citealp{ppdiag}; only simulation) and \texttt{hiddenMarkov} \citep{HiddenMarkov}.

\subsection{Uncertainty quantification, forecasting, and missing data}

Under standard regularity conditions, the maximum likelihood estimators are asymptotically normally distributed with $ \Hat{\boldsymbol{\theta}} \sim \mathcal{N}(\boldsymbol{\theta}, \mathbf{I}^{-1}), $
where $\mathbf{I}$ is the Fisher information matrix \citep{bickel1998, cappe2005}. In practice, one typically uses the observed Fisher information $\mathbf{H}$ as a plug-in estimator for $\mathbf{I}$ as it can easily be obtained from any standard numerical optimiser. 
Based on this, (approximate) standard errors and confidence intervals (CIs) for parameter estimates can be calculated. 
Alternatively, CIs can be constructed using the parametric bootstrap, but this is typically much more computer-intensive \citep{zucchini2016}. 

Out-of-sample predictions and forecasts can be calculated based on the (scaled) forward algorithm. 
Specifically, from Algorithm~\ref{algo:scaled} we obtain the scaled forward variable $\boldsymbol{\phi}_T$ at the end of the observation sequence, from which state predictions for a given time horizon $h$ can be calculated as
$$ \text{Pr}(S_{t_T+h} = i \mid x_{t_1}, \ldots, x_{t_T}) = \boldsymbol{\phi}_T \boldsymbol{\Omega}^{(h)} \mathbf{e}_i, $$
where, in a slight abuse of notation, $\boldsymbol{\Omega}^{(h)} = \boldsymbol{\Gamma}^h$ for discrete-time models (with $h \in \mathbb{N}$) and $\boldsymbol{\Omega}^{(h)} = \boldsymbol{\Gamma}(h)$ for continuous-time models  (with $h \in \mathbb{R}_{>0}$), and where $\mathbf{e}_i$ is a column vector of zeros with a one in the $i$th position only. 
Similarly, the forecast distribution for $x_{t_T+h}$ is obtained as
$$ f(x_{t_T+h} \mid x_{t_1}, \ldots, x_{t_T}) = \boldsymbol{\phi}_T \boldsymbol{\Omega}^{(h)}  \mathbf{P}(x_{t_T+h}) \boldsymbol{1}, $$
which we make use of in our case study on SSMs in Section~\ref{s:SSM}. 

In many real-world data sets, it is necessary to deal with missing observations. 
Particularly when modelling data in discrete time, these missing observations need to be accounted for in the likelihood evaluation to maintain a regular sampling grid with equally spaced time intervals between observations.  
If we can assume that the observations are missing at random, meaning that the missingness is ignorable, the likelihood expression in (\ref{eq:llk}) can easily be adjusted in a discrete-time setting by replacing the diagonal matrix $\mathbf{P}(x_{t_\tau})$ for a missing observation $x_{t_\tau}$ with the identity matrix. 
For continuous-time models, in contrast, observations missing at random can be ignored, simply leading to longer time intervals between consecutive observations accounted for by the model formulation. 
If the missingness is informative, however, more elaborate methods are required, depending on the cause of the missing data (\citealp{little2002}; for HMMs, see, e.g., \citealp{pandolfi2023} and references therein). 
To name but one example, a prevalent issue in longitudinal data is missing observations due to informative dropout, for which different methods have been developed --- especially for HMMs --- to avoid possible biases in data analyses (e.g., \citealp{spagnoli2011hidden, maruotti2015handling, bartolucci2015discrete, bartolucci2019shared}).

\section{Case studies}
\label{s:appl}

In this chapter, we provide five relatively simple real-data examples to illustrate the usage of the different model classes discussed above. 
These case studies were all implemented using \texttt{LaMa}, with the associated R code provided at \url{https://github.com/janoleko/LatentMarkov}, allowing readers to fully replicate each of the examples.
For each case study, we provide both the vanilla R code to specify the negative log-likelihood function as well as a high-performance version using \texttt{LaMa} functions as building blocks.

\subsection{HMMs}
\label{s:appl:HMMs}

We illustrate discrete-time HMMs by analysing the movement track of an elephant from Etosha National Park (ID 2; \citealp{tsalyuk2019temporal}). 
The data are available online (cf.\ the R code provided), comprising 5666 hourly GPS measurements of the elephant's locations collected between October 2008 and June 2009. 
We regard the observed movement metrics as noisy measurements of the underlying behavioural state process, where the latter is typically of primary interest in movement ecology. Furthermore, it is natural to assume the state process to be discrete-valued, corresponding to distinct behaviours of the animal such as resting, foraging, or travelling \citep{beumer2020}. 
We thus model the data using a discrete-time HMM, where the observed time series of movement metrics is assumed to be driven by a latent discrete-time Markov chain roughly corresponding to the behavioural modes.
Since the observations are made at equidistant, non-informative time points, the transition probabilities can be interpreted with respect to the fixed sampling unit of one hour.

\begin{figure}
    \centering
    \includegraphics[width=1\textwidth]{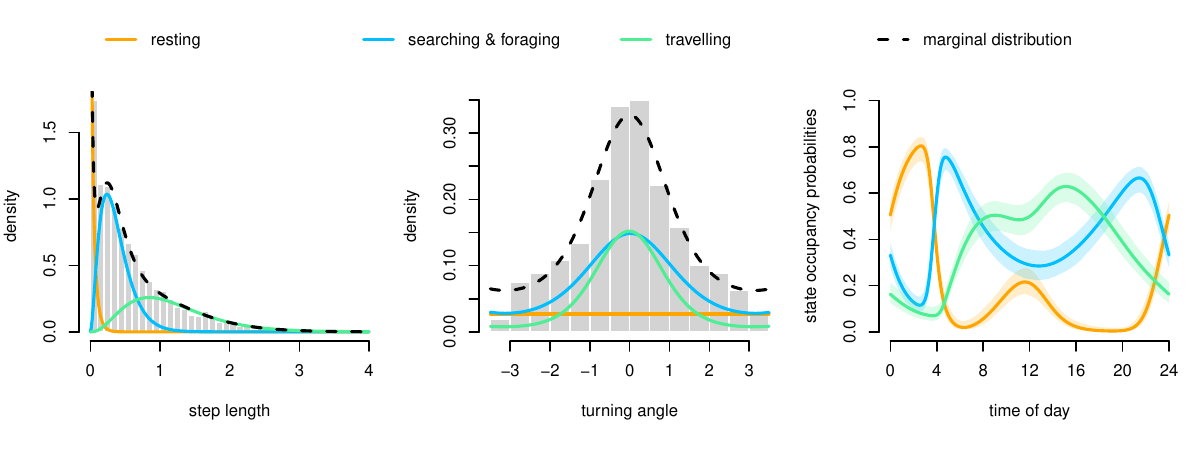}
    \caption{HMM fitted to elephant movement data --- state-dependent distributions for step length (left panel) and turning angle (middle panel) as well as the state occupancy probabilities as a function of time of day (right panel).}
    \label{fig:elephant}
\end{figure}

Instead of directly modelling the GPS locations, the observations were first transformed into a bivariate time series of step lengths and turning angles \citep{michelot2016movehmm}. 
We assume {contemporaneous conditional independence} of the observed variables, given the underlying behavioural state, such that the density of the joint state-dependent distribution is obtained as the product of the density of a gamma distribution for the step lengths and the density of a von Mises distribution for the turning angles.
To learn something about the mechanisms of the behavioural state dynamics, we further model the transition probabilities as functions of the time of day via linear predictors of the form
$$
\eta_{ij}^{(t)} = \beta^{(ij)}_0 + \beta_1^{(ij)} \sin \Bigl( \frac{2 \pi \text{time}_t}{24} \Bigr) + \beta_2^{(ij)} \cos \Bigl( \frac{2 \pi \text{time}_t}{24} \Bigr), \quad i, j = 1, \dotsc, N, \quad i \neq j, 
$$
where $\eta_{ii}^{(t)}=0$ and where the $i$th row of the t.p.m.\ is computed from $\eta_{ij}^{(t)}, j = 1,\dotsc,N$ by applying the inverse multinomial logistic link \citep{patterson2017statistical}. 

The fitted state-dependent distributions, displayed in the left and middle panel in Figure \ref{fig:elephant}, allow for an interpretation of the states as proxies for resting, searching \& foraging, and travelling behaviour, respectively. The right panel in 
Figure \ref{fig:elephant} shows the model-implied state occupancy probabilities as a function of time, obtained as described in \citet{koslik2023inference}. 
We find that resting is the dominant behaviour at nighttime, that searching \& foraging occurs primarily in the early morning as well as the late evening hours, and that the probability of exhibiting travelling behaviour peaks around noon.

\subsection{ctHMMs}

Continuous-time HMMs are required if the data are collected irregularly in time, preventing the interpretation of the transition probabilities with respect to a fixed sampling unit. Such data often arise in medical contexts, where new patient data typically arrives whenever the doctor is visited. 
We consider precisely such a data set, where the observations made over time concern the forced expiratory volumes in one second (FEV) measured for lung transplant recipients, expressed as a percentage of the baseline value, from six months onwards after their transplant \citep{jackson2002hidden}. This data set is available in the R package \texttt{msm} \citep{jackson2011msm} and comprises 5896 measurements from 203 patients, of whom 96 died during the observation period. 
Similar to \citet{jackson2002hidden}, we model the patients' underlying disease progression as a 3-state continuous-time Markov chain, where the first state corresponds to ``healthy", the second state combines two stages of \textit{obliterative bronchiolitis}, a chronic condition causing irreversibly airway obstruction and reduced lung function, and the third (absorbing) state indicates death. 

\begin{figure}
    \centering
    \includegraphics[width = 0.9\textwidth]{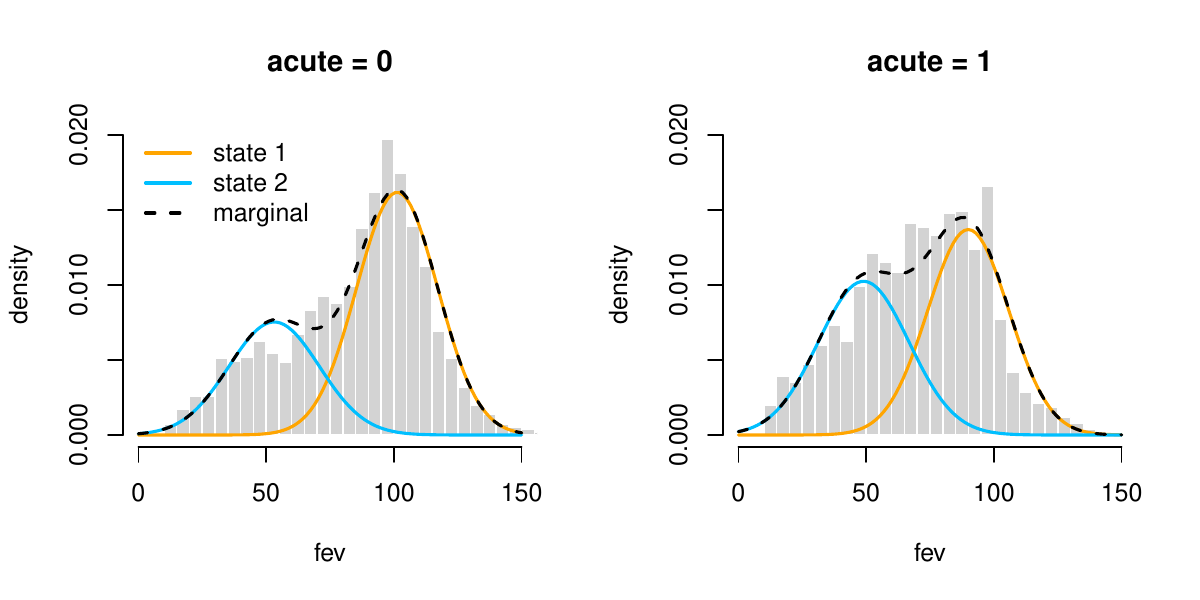}
    \caption{ctHMM fitted to FEV measurements --- state-dependent distributions in the healthy (orange) and diseased (light blue) state, without an acute infection (left panel) and with an acute infection (right panel) in the last 14 days before measurement.}
    \label{fig:enter-label}
\end{figure}

The irreversibility of the bronchiolitis obliterans syndrome (BOS) dictates a clear state transition structure: patients can only transition from the healthy state to the diseased state or death and from the diseased state to death. Thus, we specify the generator matrix as
$$
\mathbf{Q} = \begin{pmatrix}
    -(q_{12}+q_{13}) & \; q_{12} \; & \; q_{13} \; \\
    0 & -q_{23} & q_{23} \\
    0 & 0 & 0
\end{pmatrix},
$$
fixing transition rates associated with impossible transitions at zero. As a consequence, the last row of $\mathbf{Q}$ consists of only zeros, as its last entry needs to be equal to the sum of the transition rates to state 1 and state 2.
We specify the state-dependent distribution of the observed FEV variable as
$$
\text{FEV}_{t_{\tau}} \mid S_{t_{\tau}}=j \sim     \mathcal{N}\bigl( \mu^{(t_{\tau})}_j, \sigma_j^2\bigr) $$
for $j = 1, 2$, i.e.\ alive patients with an actual FEV measurement, while death observed at time $t_{\tau}$ is tied to $S_{t_{\tau}}=3$ via specifying 
$\mathbf{P}(x_{t_\tau}) = \text{diag} (0,0,1)$.
We further model the conditional mean of FEV as $\mu^{({t_\tau})}_j = \beta_{j0}+\beta_{j1} \text{acute}_{t_\tau}$, where $\text{acute}_{t_\tau}$ is a dummy variable indicating whether the patient suffered from an acute infection during the 14 days prior to time ${t_\tau}$.

The fitted model as displayed in Figure \ref{fig:enter-label} clearly distinguishes a healthy and a diseased state and shows a significant impact of an acute infection in the last 14 days on the forced expiratory volume (healthy state: $\hat{\beta}_{11} = -11.3 \; (-12.61, -9.99)$; diseased state: $\hat{\beta}_{21} = -4.12 \; (-5.90, -2.33)$, 95\% CIs in brackets; these effects can be interpreted as percentage reductions of the forced expiratory volume compared to the baseline measurement). From the estimated generator matrix, we can derive the average time it takes patients to transition from the healthy state to BOS as 4.96 years, and the average time for a transition from BOS to death as 3.5 years (compared to an average of 26.57 years for a direct transition from healthy to death, i.e.\ without developing BOS).

\subsection{SSMs}
\label{s:SSM}

Stochastic volatility modelling is a well-known area of application of  discrete-time SSMs. Despite their relative simplicity --- in its most basic form, a basic SSM for market volatility comprises only three parameters --- these models capture most of the stylised facts commonly found in financial share returns, such as volatility clustering and the heavy-tailed distribution of returns \citep{cont2001empirical}. The standard discrete-time stochastic volatility model for log returns $y_t = \log(\text{Close}_t / \text{Close}_{t-1})$ can be written as
$$
y_t = \mu + \beta \epsilon_t \exp(g_t / 2), \qquad g_t = \phi g_{t-1} + \sigma \eta_t, \qquad t = 1, \dotsc, T,
$$
where $\epsilon_t$ and $\eta_t$ are independent and identically standard normally distributed innovations. The parameters of interest are the long-term average return $\mu$, the baseline standard deviation $\beta$ when $\{g_t\}$ is in equilibrium, the autoregression parameter $\phi$ with $\lvert \phi \rvert < 1$, and $\sigma$ governing the dispersion of the underlying volatility \citep{shephard1996statistical}.

In our illustrating example, we model Bitcoin returns from September 2014 until June 2024,  which we downloaded from yahoo.com using the R package \texttt{fHMM} (\citealp{fHMM}; cf.\ the R code provided). 
A similar analysis was conducted by \citet{pennoni2022exploring}, who modelled several cryptocurrencies using a multivariate Gaussian distribution and a six-state HMM instead of an SSM. 
While the parameter $\mu$ is typically set to zero, we chose to estimate this parameter freely given the strong increase of the Bitcoin value within the observation period, suggesting a slightly positive expected return. The model can be fitted via direct numerical approximate maximum likelihood estimation by choosing the number of states of the approximating HMM and the essential range of $\{g_t\}$ for the numerical integration as detailed in Section \ref{s:inference}. 
\begin{figure}
    \centering
    \includegraphics[width=0.9\textwidth]{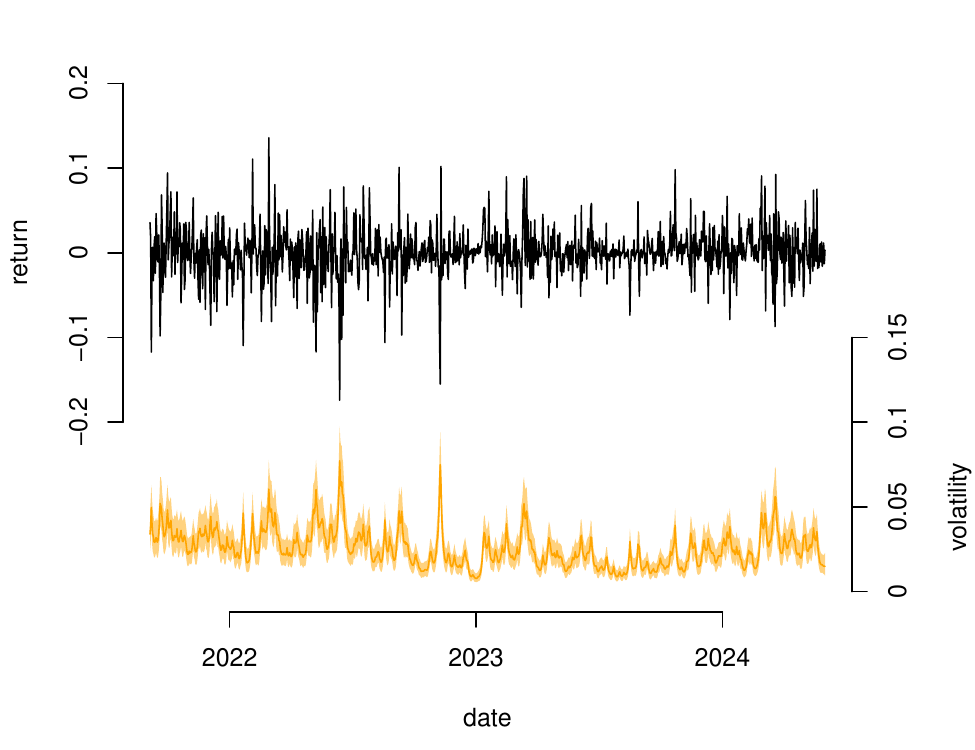}
    \caption{SSM fitted to Bitcoin returns --- log returns from the subset of the data covering the time from July 2021 to March 2024 (black), complemented with the decoded volatility $\exp(g_t / 2) = \sigma_t / \beta$ $\pm$ one standard deviation (orange).}
    \label{fig:bitcoin}
\end{figure}
In this simple model, both the transition density of the latent volatility process and the state-dependent distributions are Gaussian --- which could easily be extended to other distributions, e.g., a state-dependent $t$-distribution. Choosing $m=200$ states and a range of $(-3.5, 3.5)$ for the approximation, we arrive at estimates $\Hat{\beta} = 0.026$ (0.024, 0.028), $\Hat{\phi} = 0.888$ (0.852, 0.916), $\Hat{\sigma} = 0.554$ (0.483, 0.636), and $\Hat{\mu} = 0.0012$ (0.0006, 0.0019). The most probable latent volatility sequence, derived using the Viterbi algorithm \citep{viterbi1967, forney1973}, as well as the original time series are shown in Figure \ref{fig:bitcoin}. 

Using standard HMM machinery \citep[][Chapter 5]
{zucchini2016}, we can arrive at the one-day-ahead forecast distribution for the daily return, allowing us to calculate the value at risk (VaR) --- a quantity of interest in risk management --- which we consider here to evaluate our model.  Specifically, we re-fitted the model to the return data only until 31 December 2020.
Subsequently, for each day from 1 January 2021 until 1 June 2024, we calculate the 1\% VaR based on the one-day-ahead forecast distribution. In the validation period of length 1201 days, the model-implied VaR was exceeded 14 times, corresponding to a relative frequency of $0.0112$, which closely matches the target level.

\subsection{ctSSMs}

To illustrate the application of ctSSMs, we consider data on sports performances. 
In this area, relevant performance measures such as the outcomes of free throws in basketball are often irregularly sampled over time, thus requiring a continuous-time model formulation. 
In addition, there is interest in capturing possible ``hot hand'' effects, according to which athletes can experience elevated success probabilities over extended periods of time \citep{bar2006twenty}. 
This phenomenon has received much attention in both the sports sciences and psychology. 
Yet, statistical evidence for the existence of such an effect is still mixed, with some authors attributing the belief in the hot hand to a tendency of people to overinterpret streaks of successes and failures \citep{gilovich1985hot,bar1991perception,bar2006twenty}.  
Conceptually, a possible hot hand effect naturally translates to a latent state process corresponding to the current form (``hotness'') of a player or team, and as there is no intuitive reason for the form to take on only finitely many values, it seems sensible to assume the state process to be continuous-valued \citep{stone2012measurement,otting2020hot}. 
The class of ctSSMs hence is a natural framework for investigating serial correlation in sports performances \citep{mews2023continuous}.

Here we provide an example analysis of a possible hot hand effect based on ctSSMs, considering sequential seven-metre throws in handball. We extracted corresponding sequences from a play-by-play data set for the German Handball Bundesliga 2023/2024 which, after preprocessing, resulted in a total of 1131 seven-metre throws performed by 46 players in 205 matches. 
The data are available online together with the provided R code.
We model the outcome of the $k$th seven-metre throw as a Bernoulli trial with success probability $\pi_{t_k}$, where $\text{logit}(\pi_{t_k}) = \beta_0 + S_{t_k}$ and where the unobserved state process $\{S_t\}$ is modelled as an OU process as described in Section~\ref{ss:modelFormulation}, here representing the underlying ``hotness" of the player. Given the underlying state, the observation is assumed to be independent of previous throws, with the state process inducing serial correlation (if there is any). To avoid identifiability issues, we set the mean of the OU process to $\mu = 0$.

\begin{figure}[!htb]
    \centering
    \includegraphics[width=1\textwidth]{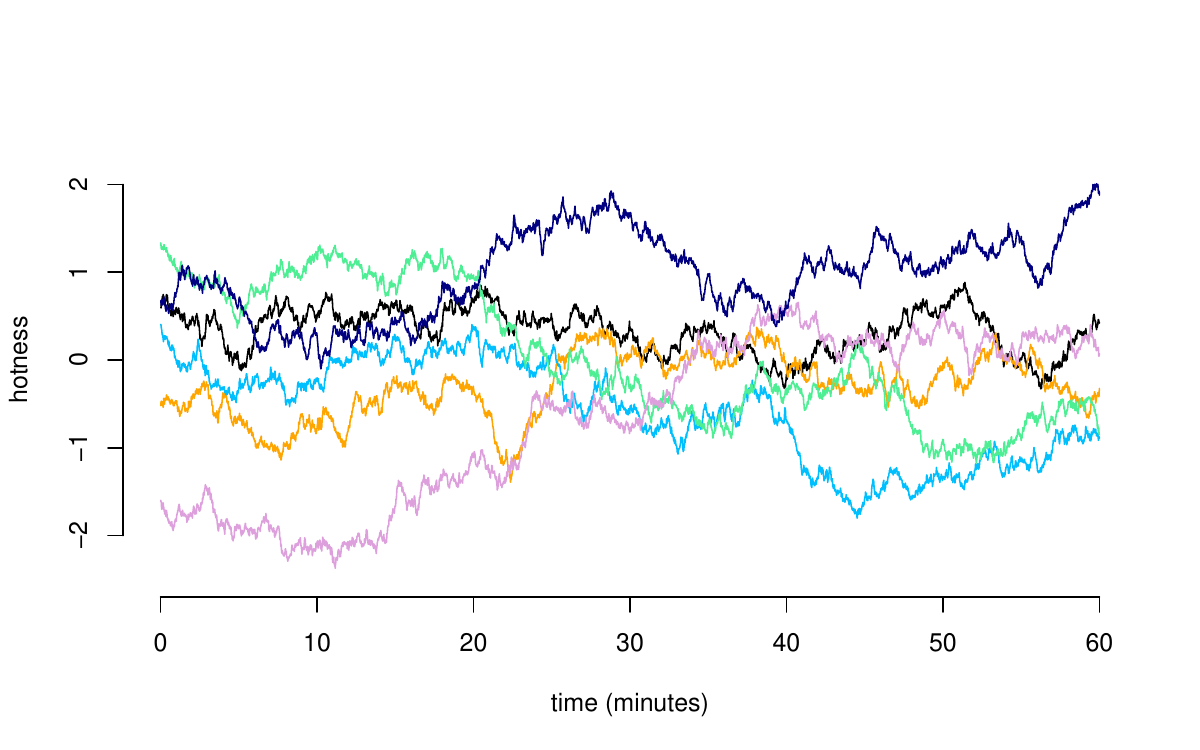}
    \caption{Six sample path realisations of the OU process underlying the ctSSM fitted to the seven-metre throw data. The graphs were obtained by the Euler-Maruyama scheme with step length 1/60 and an initial value drawn from the stationary distribution of the OU process.}
    \label{fig:simOU}
\end{figure}
The model can again be fitted by direct numerical approximate maximum likelihood estimation, choosing the number of states of the approximating HMM and the essential range of the OU process as explained in Section \ref{s:inference}. With $m = 250$ states and a range of $(-3.5, 3.5)$ for the state process, the fitted model yields the estimates $\Hat{\theta} = 0.019 \; (0.004, 0.099)$, $\Hat{\sigma} = 0.21 \; (0.087, 0.504)$, and $\Hat{\beta} = 1.519 \; (1.305, 1.732)$, corresponding to an average scoring probability of $0.8204$ when the state process is in equilibrium. The autocorrelation of the OU process is proportional to $\exp(-\theta \Delta t)$, where $\Delta t$ denotes the length of the time interval considered. Plugging in $\Hat{\theta} = 0.019$ yields an autocorrelation of 91\% after five minutes and still 32\% after 60 minutes, where the latter is the length of an entire match. 
Because of this high temporal persistence, the estimated model suggests the presence of a ``hot hand" effect in handball seven-metre throws. 
However, it needs to be stressed that the confidence interval for $\theta$ is fairly wide and that the model formulation may be too simplistic to capture the complexity of the team sport handball --- for more reliable and detailed inference, the model could be extended by adding covariates to the state-dependent process to control for individual variation of the players or team effects. 
To illustrate the behaviour of the OU process of the fitted ctSSM, Figure \ref{fig:simOU} shows six sample paths obtained from the fitted model, clearly indicating persistence in the state process, i.e.\ a varying ``hotness''.

\subsection{MM(M)PPs}

\begin{figure}[!htb]
    \centering
    \includegraphics[width=0.9\textwidth]{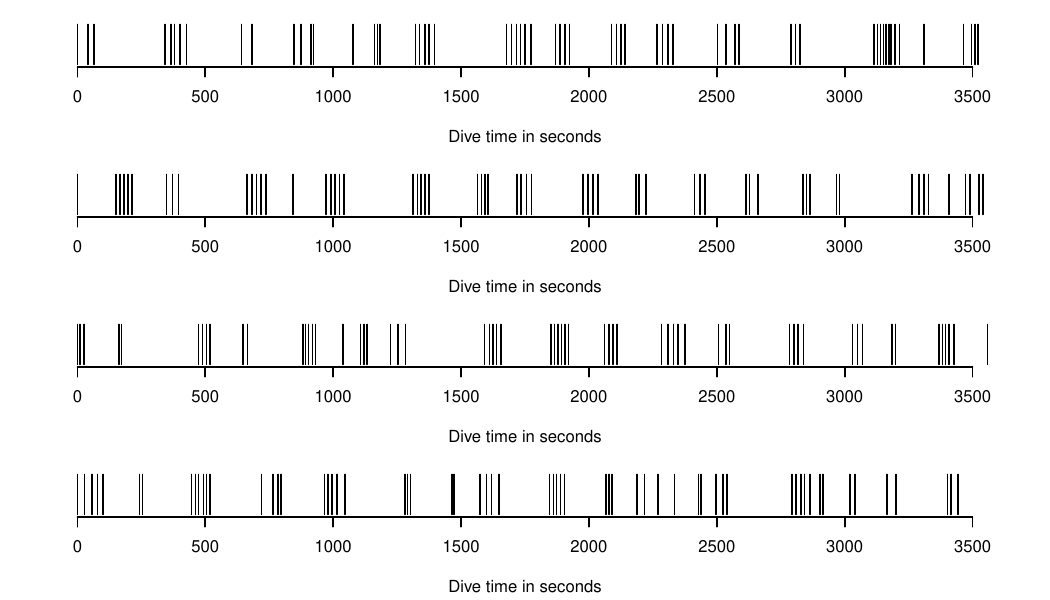}
    \caption{Surfacing times of four minke whales.}
    \label{fig:mmpp}
\end{figure}

To illustrate a situation in which the observation times themselves depend on the state of the underlying process, we consider sequences of surfacing times of four minke whales \textit{(Balaenoptera acutorostrata)}, displayed in Figure \ref{fig:mmpp}. 
The data and R code are provided online.
Adequately modelling such stochastic surfacing times is of paramount importance in distance sampling surveys for estimating marine mammal abundance, as estimates of the population size need to account for individuals not being available to be sighted (and hence counted) due to being submerged 
\citep{borchers2013using}. 

Surfacing times of marine mammals have sometimes been modelled using simple homogeneous Poisson processes \citep{skaug1999hazard, okamura2003abundance}. Still, for the minke whale data, it is evident that these whales alternate between sequences of several short shallow dives and much longer foraging dives, leading to a clustering of the surfacing events --- these data hence suggest the use of MMPPs (cf.\ \citealp{langrock2013markov,borchers2015double}).

We model these data using 2-state MMPPs (without marks), fitting separate models to each individual to account for potential heterogeneity. All four fitted models comprise one state that is associated with shallow dives (average times between surfacings of 27--34 seconds; average sojourn times in the state of 65--95 seconds) and another state that is associated with deep foraging dives (no surfacings, i.e.\ state-dependent Poisson arrival rates of approximately zero; average times in the state of 59--78 seconds). 
Such an MMPP for the whales' sighting availability can be embedded in distance sampling models where the probability of detection, given availability, depends on the distance between whale and observer \citep{langrock2013markov}. 
In this setting, additional marks could be incorporated in the stochastic availability process to model whether or not a surfacing event is accompanied by a visible spout (i.e.\ the whale expelling air through the blowhole), which would increase the chance of the surfacing being detected --- this would lead to an MMMPP, but we do not have corresponding data available to illustrate this additional extension (but see, e.g., \citealp{mews2023markov},  for an example in the medical context).

\section{Discussion}
\label{s:discussion}

This paper attempts to provide a unified view on several distinct, but closely related model classes --- discrete- and continuous-time HMMs (including Markov-switching regression models), discrete- and continuous-time SSMs, and Markov-modulated (marked) Poisson processes --- a) by identifying key commonalities shared by these models concerning the associated inferential tools, b) by discussing how to choose between these models in practice, and c) by providing software, the \texttt{LaMa} package, for simple and fast implementation of the most important part of the model fitting, namely the likelihood evaluation. 
We suggest to summarise these model classes under the label ``latent Markov models'', despite the fact that this term has previously been used in a slightly narrower sense \citep{bartolucci2012latent,bartolucci2014latent,bacci2014}.

Our optimistic vision is that this unified view on several closely related model classes will allow for the conceptualisation and development of corresponding lectures, textbooks and software much like this has been accomplished for generalised linear models (GLMs) following the seminal paper by \citet{nelder1972generalized}. 
Compared to GLMs, the main challenges lie in the higher numerical instability of the more complex latent Markov models and in the relatively large number of researcher degrees of freedom due to the numerous possible ways to extend basic models by incorporating covariates, random effects, and additional dependence structure.   
Despite these challenges, we see it as important to regard the various special cases of latent Markov models as members of a larger family of models. 
Such a unifying view will provide practitioners with access to a large suite of statistical approaches for analysing sequential data, allowing for a systematic selection of the most adequate model formulation based on the characteristics of the data at hand. 

\section*{Supplementary materials}
The data and code for fully reproducing all case studies can be found at \url{https://github.com/janoleko/LatentMarkov}.


\section*{Acknowledgements}

The authors would like to thank Rouven Michels for providing the handball data analysed in Section 4.4 and Andrea Langrock for helping us to produce a nicer version of Figure 1.

\section*{Declaration of conflicting interests}
The authors have no potential conflicts of interest with respect to the research, authorship and/or publication of this article to declare.

\section*{Funding}

This research was partly funded by the German Research Foundation (DFG) as part of the SFB TRR 212 (NC$^3$); project numbers 316099922 and 396782756.

\bibliography{references}

\end{document}